\documentclass[onecolumn]{aastex63}

\newcommand{\dom}{\Delta\Omega}
\newcommand{\kms}{\mathrm{km\,s^{-1}}}
\newcommand{\norm}{\mathcal{N}}
\newcommand{\vmac}{\zeta}
\newcommand{\vtide}{v_\mathrm{tidal}}
\newcommand{\mrg}{M_\star}
\newcommand{\mc}{M_\bullet}
\newcommand{\rrg}{R_\star}
\newcommand{\logu}{{\mathcal{U}_{\ln}}}
\renewcommand{\u}{\mathcal{U}}
\newcommand{\leff}{\lambda_\mathrm{eff}}

\usepackage{amsmath}
\usepackage{comment}

\shorttitle{Tidal RVs in Circular and Synchronized Binaries}
\shortauthors{Masuda and Hirano}
\graphicspath{{./}{figures/}}

\begin{document}

\title{ 
Tidal Effects on the Radial Velocities of V723 Mon: Additional Evidence for a Dark 
$3\,M_\odot$ Companion
}

\correspondingauthor{Kento Masuda}
\email{kmasuda@ess.sci.osaka-u.ac.jp}

\author[0000-0003-1298-9699]{Kento Masuda}
\affiliation{Department of Earth and Space Science, Osaka University, Osaka 560-0043, Japan}

\author[0000-0003-3618-7535]{Teruyuki Hirano}
\affiliation{Astrobiology Center, NINS, 2-21-1 Osawa, Mitaka, Tokyo 181-8588, Japan}
\affiliation{National Astronomical Observatory of Japan, NINS, 2-21-1 Osawa, Mitaka, Tokyo 181-8588, Japan}




\begin{abstract}

\citet{2021arXiv210102212J} identified a dark $\approx 3\,M_\odot$ companion on a 
nearly edge-on $\approx 60\,\mathrm{day}$
orbit around the red giant star V723 Monoceros as a black hole candidate in the mass gap. This scenario was shown to explain most of the data presented by \citet{2021arXiv210102212J}, except for periodic radial velocity (RV) residuals from the circular Keplerian model. Here we show that the RV residuals are explained by orbital phase-dependent distortion of the absorption line profile associated with changing visible fractions of the approaching and receding sides of the red giant star, whose surface is tidally deformed by and rotating synchronously with the dark companion. Our RV model constrains the companion mass $M_\bullet = 2.95\pm0.17\,M_\odot$ and orbital inclination $i=82.9^{+7.0}_{-3.3}\,\mathrm{deg}$ (medians and 68.3\% highest density intervals of the marginal posteriors) adopting the radius of the red giant $24.0\pm0.9\,R_\odot$ as constrained from its SED and distance. The analysis provides independent support for the companion mass from ellipsoidal variations and the limits on the companion's luminosity from the absence of eclipses both derived by \citet{2021arXiv210102212J}. We also show that a common scheme to evaluate the tidal RV signal as the flux-weighted mean of the surface velocity field can significantly underestimate its amplitude for RVs measured with a cross-correlation technique, and present a modified prescription that directly models the distorted line profile and its effects on the measured RVs. The formulation will be useful for estimating the component masses and inclinations in other similar binaries. 

\end{abstract}



\section{Introduction} \label{sec:intro}

\citet{2021arXiv210102212J} reported that the red giant (RG) star V723 Mon is orbited by a dark $\approx 3\,M_\odot$ companion on a nearly circular and edge-on orbit with the period of $P \approx 60\,\mathrm{days}$. 
If the companion is a single compact object, it is the nearest known black hole that falls within the ``mass gap" \citep{1998ApJ...499..367B, 2011ApJ...741..103F}, perhaps along with another similar system discovered by \citet{2019Sci...366..637T}. 
This scenario successfully explained most of the data presented in  \citet{2021arXiv210102212J}, but there remained one signal yet to be explained: the 
radial velocity (RV)
time series of V723 Mon exhibit periodic residuals from the signal due to Keplerian orbital motion. The residual RVs have the period of $P/2$ when the binary orbit is assumed to be circular, and $P/3$ when the orbital eccentricity is fitted (in which case a small eccentricity eliminates the $P/2$ component). 
The periodic residuals motivated \citet{2012AN....333..663S} to propose the presence of a third body, although the triple scenario would suffer from fine tuning due to dynamical stability \citep{2014Obs...134..109G, 2021arXiv210102212J} and the residual signal appears to be too large to arise from a dynamically stable triple configuration \citep{2020ApJ...890..112H}.

\begin{figure*}
    \centering
    \epsscale{1.1}
    \plottwo{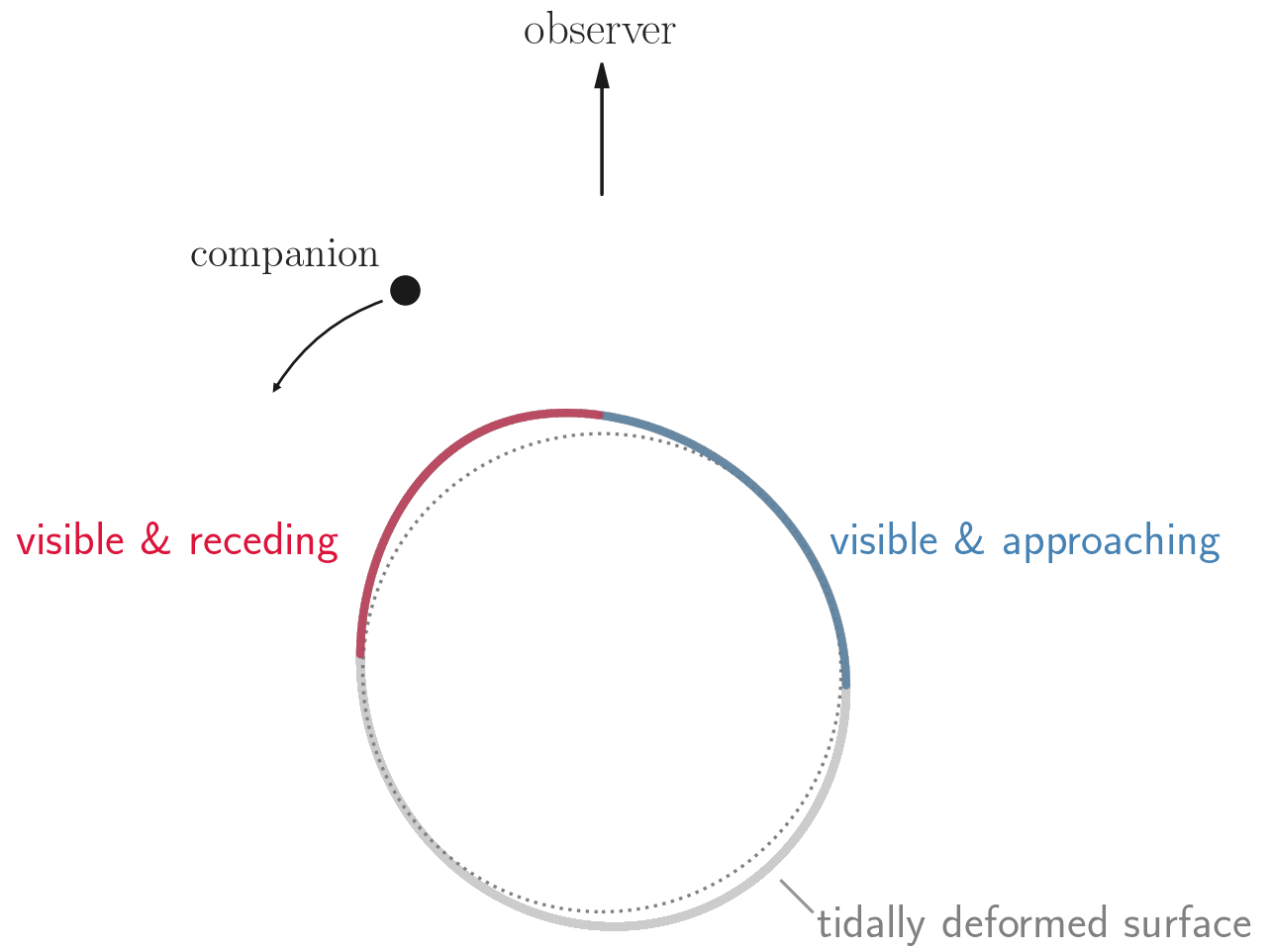}{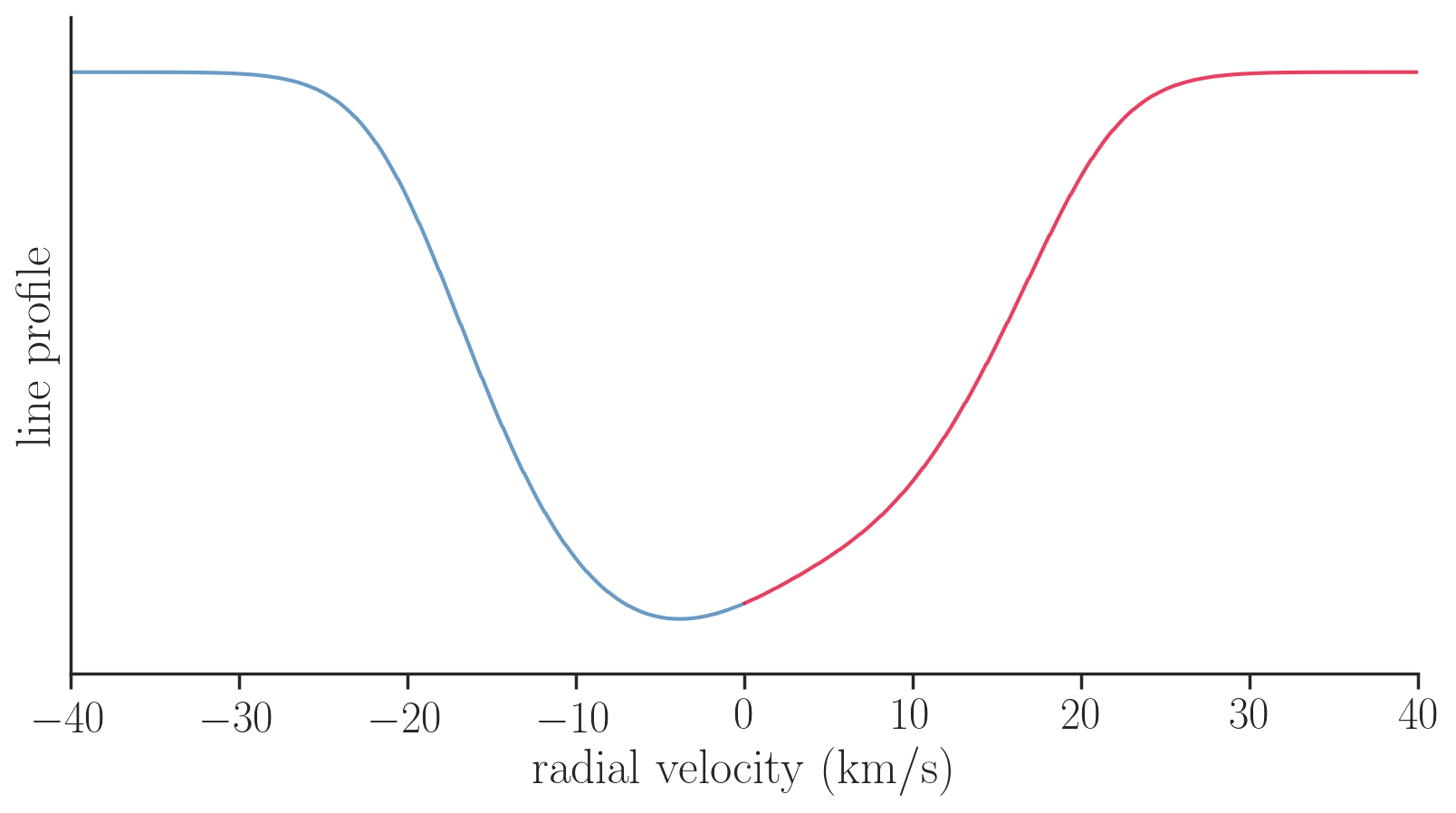}
    \caption{Schematic illustration of the tidal effect on the absorption line profile. At this orbital phase the red giant star appears to be blue shifted and exhibits a negative anomalous radial velocity. {\it (Left)---} The thick gray line shows the equator of the red giant deformed by the companion. The companions' orbit and stellar equator are both assumed to be edge-on as seen from the observer. Note that the companion's orbit and red giant radius are not to scale. The dotted line shows a circle to emphasize the asymmetric deformation of the red giant.
    {\it (Right)---} The absorption line profile corresponding to the configuration shown in the left panel, computed in a manner as described in Section \ref{ssec:method_rv}.}
    \label{fig:schematic}
\end{figure*}

Here we show that the periodic RV residuals originate from tidal deformation of the red giant whose rotation is synchronized with the binary orbit, as was argued to be a plausible scenario by \citet{2021arXiv210102212J}.
The deformation causes orbital phase-dependent modulation of the fractions of the visible RG surfaces that are moving toward and away from us (Figure \ref{fig:schematic}). The imbalance causes asymmetric distortion of the absorption line profile and produces RV anomalies.
This picture is consistent with the phase-dependent variations of the projected rotation velocity $v\sin i$ \citep{2014Obs...134..109G, 2021arXiv210102212J} and also explains the different RV anomalies around the orbital phases at 0 and 0.5 
\citep[as seen in the middle panel of Figure 3 in][]{2021arXiv210102212J} 
because the RG surface becomes more elongated at the near side of the companion than the far side due to strong tides (Figure \ref{fig:schematic} left).
Furthermore, 
the shape of the RV curve is sensitive to the orbital inclination: 
the RV anomaly is small for nearly pole-on systems, while a sharp anomaly arises around the conjunction where the companion is in front for nearly edge-on systems \citep[e.g., Figure 7 of][]{2008ApJ...681..562E}. 
All these features make the ``tidal RVs" sensitive to the masses of the binary components.

Significant tidal deformation also manifests as ellipsoidal variations in the photometric light curve, which was used by \citet{2021arXiv210102212J} to precisely constrain the component masses and orbital inclination in combination with the precise binary mass function from RVs and the prior on the RG radius. 
The inferred orbit is very close to edge-on, and this is also consistent with the eclipses of the Balmer emission observed when the companion is supposed to be behind the red giant. That said, the light curve model includes wavelength-dependent dilution (the ``veiling" component) whose physical origin is yet to be understood, and any inaccurate assumption on such non-stellar flux could become a source of systematic errors in the mass/inclination measurements with ellipsoidal variations \citep{2012ApJ...757...36K}.
Similarly, it is not yet clear how and where the Balmer emission is produced \citep{2021arXiv210102212J}. 
The mass measured with ellipsoidal variations could also be biased by quasi-periodic photometric modulations associated with active regions (spots) on the RG surface, whose contribution is difficult to evaluate a priori.
Therefore a constraint on the component masses that does not rely on these signals would be valuable. Such a technique would also allow for mass measurements in a larger number of similar systems for which photometric data are not available and/or ellipsoidal variations are significantly contaminated by other signals.

To our knowledge, this tidal RV signal was first discussed by \citet{1941PNAS...27..168S} as a source of spurious eccentricity in spectroscopic binaries, and a numerical scheme for more accurate modeling was given by \citet{1976ApJ...203..182W} \citep[see also other references therein, including][]{1959cbs..book.....K}. 
Although this signal has been clearly detected only in a handful of systems
(but see 
Figure 4 of \citet{1989A&A...218..152H} and Figure 10 of 
\citet{2008ApJ...681..562E} for some notable examples), the signal has still been recognized as a correction that needs to be taken into account in interpreting the RV data of tidally-locked binaries \citep[e.g.][]{1986ApJ...308..110M, 1986AJ.....91..125K}, and is also implemented in the widely used ELC code \citep{2000A&A...364..265O}. 
We find, however, that the scheme adopted in these previous works to evaluate the tidal RV signal as the flux-weighted mean of the surface velocity field, following the prescriptions in the earlier works \citep{1941PNAS...27..168S, 1976ApJ...203..182W}, 
significantly underestimates its amplitude in the RVs of V723 Mon measured with a cross correlation technique (Figure \ref{fig:rvmethod}). 
This is because the tidal effects work to {\it distort} the stellar lines, 
rather than to shift them, and the peak (trough) of the distorted, asymmetric profile as probed by cross correlation is different from its centroid as evaluated by computing the flux-weighted mean (see Figure \ref{fig:schematic} right). In other words, the 
RV derived from the flux-weighted mean of the Doppler shifted profiles 
is not the same as the flux-weighted mean of the Doppler shifts.
In this paper, we present a formulation that explicitly models the line profile and cross-correlation procedure, and show that it is indeed crucial for modeling the high signal-to-noise anomaly in the RVs of V723 Mon.\footnote{We note that the need for such a treatment was also recognized in some earlier works including \citet{1985A&A...152...25V}, \citet{1989A&A...218..152H}, and \citet{1993ASPC...38..127H}, although the scheme was not used to fit the actual data.}

The remainder of this paper is organized as follows.
In Section \ref{sec:model}, we present our model for the tidal RV signal in a circularized and synchronized binary. In Section \ref{sec:results}, we show that the model quantitatively reproduces the RV residuals observed in V723 Mon, and derive constraints on the system parameters based on the RV data. We find independent support for a $3\,M_\odot$ companion in a nearly edge-on orbit, and have eliminated the need for a third body to explain the RV residuals. In Section \ref{sec:summary} we summarize the results and discuss future prospects.

\section{The Model}\label{sec:model}

We assume that the binary orbit has been circularized, and rotation of the red giant has been synchronized with the orbital motion (i.e. rotation period and axis are the same as the orbital ones). 
In this case, the red giant is static in the rotating frame, and each surface element moves on a circular orbit. 
We compute the geometric shape and flux distribution over the deformed surface of the red giant following a standard procedure (Section \ref{ssec:method_flux}). We then use them to model variations in the absorption line profiles as the companion and red giant rotate together, and translate the phase-dependent distortion of the line profile into the tidal RV signal $\vtide$ (Section \ref{ssec:method_rv}) --- this step is not included in the  formulation by \citet{1976ApJ...203..182W}.
The model is compared with observed RVs and the parameters are constrained via a Bayesian formalism adopting appropriate priors (Sections \ref{ssec:method_full} and \ref{ssec:method_priors}).

\subsection{Shape of the Tidally Deformed Surface and Flux Distribution}\label{ssec:method_flux}

We divide the stellar surface into $768$ pixels\footnote{This gives angular resolution of $\approx 7.3^\circ$, which corresponds to the velocity resolution of $\sim 1\,\kms$ for the red giant star of interest. This resolution is sufficient because it is smaller than the intrinsic velocity width of the absorption lines (Section \ref{ssec:method_rv}).}
with equal solid angle $\dom$ using the {\tt HEALPix}/{\tt healpy} package \citep{2005ApJ...622..759G, Zonca2019}.\footnote{\url{http://healpix.sourceforge.net}} 
For each pixel labeled by $j$, we compute:
\begin{itemize}
    \item normalized distance from the star's center $R_j/R_\star$,
    \item normalized surface gravity $g_j/g_\star$,
    \item foreshortening factor $\cos \gamma_j$, where $\gamma_j$ is the angle between the surface normal and our line of sight,
    \item angle $\delta_j$ between the 
    surface normal and radius vector,
    \item intensity $I_j(g_j,\cos\gamma_j)$ that depends on $g_j$ and $\cos\gamma_j$ through gravity- and limb-darkening, respectively,
    \item line-of-sight velocity with respect to the star's center of mass normalized by the equatorial rotation velocity, $V_j/(2\pi\rrg/P)$ 
\end{itemize}
as a function of the orbital phase, orbital/spin inclination $i$, RG mass $\mrg$, companion mass $\mc$, and semi-major axis scaled by the RG radius $a/\rrg$. The flux contribution of each pixel $\Delta F_j$ is given by
\begin{equation}
    \Delta F_j \propto I_j(g_j, \cos\gamma_j)\,\cos\gamma_j\,{R_j^2\dom \over \cos\delta_j}
\end{equation}
for $\cos \gamma_j>0$ (i.e. visible to the observer), and $0$ otherwise.
The flux change due to Doppler beaming is of order $10^{-4}$ for the rotation velocity of $\approx 20\,\kms$ and is not included. We also ignore the effects of irradiation and reflection because the companion appears to be non-luminous. A potential microlensing effect due to the compact companion eclipsing the RG star is also negligible given the large RG radius and relatively tight orbit \citep{1969ApJ...156.1013T}. 

The quantities $R_j$, $g_j$, $\cos\gamma_j$ and $\cos\delta_j$ were computed assuming that the RG surface is described by a surface of the constant Roche potential \citep{1979ApJ...234.1054W}, where $R_j$ was solved iteratively for each grid point. 
The formulation is thus similar to the one in the {\tt PHOEBE} model \citep[e.g.][]{2016ApJS..227...29P}. 
The normalizations $R_\star$ and $g_\star$ were chosen to be the values at the points on the stellar equator perpendicular to the star--companion axis. 
The intensity $I_j$ was computed adopting the quadratic limb-darkening law
$I(\cos\gamma) \propto 1 - u_1(1-\cos\gamma) - u_2(1-\cos\gamma)^2$, multiplied by 
$(g_j/g_\star)^y$ \citep{1959cbs..book.....K}.
Since the limb- and gravity-darkening coefficients $u_1$, $u_2$, and $y$ are chromatic, the values of the coefficients need to be evaluated for an appropriate wavelength band, as will be discussed in Section \ref{ssec:method_priors}.
Here the quadratic law is adopted considering a balance between the accuracy and computational cost, and the model can also be modified to incorporate more complex profiles as we try in Section \ref{ssec:results_ld}. Although the quadratic law may fail to reproduce the limb-darkening at the very edge of the stellar disk accurately, the results from the modeling in this paper was found to be insensitive to the adopted profile.

\subsection{Tidal RVs}\label{ssec:method_rv}

\begin{figure}
    \centering
    \epsscale{1.1}
    \plotone{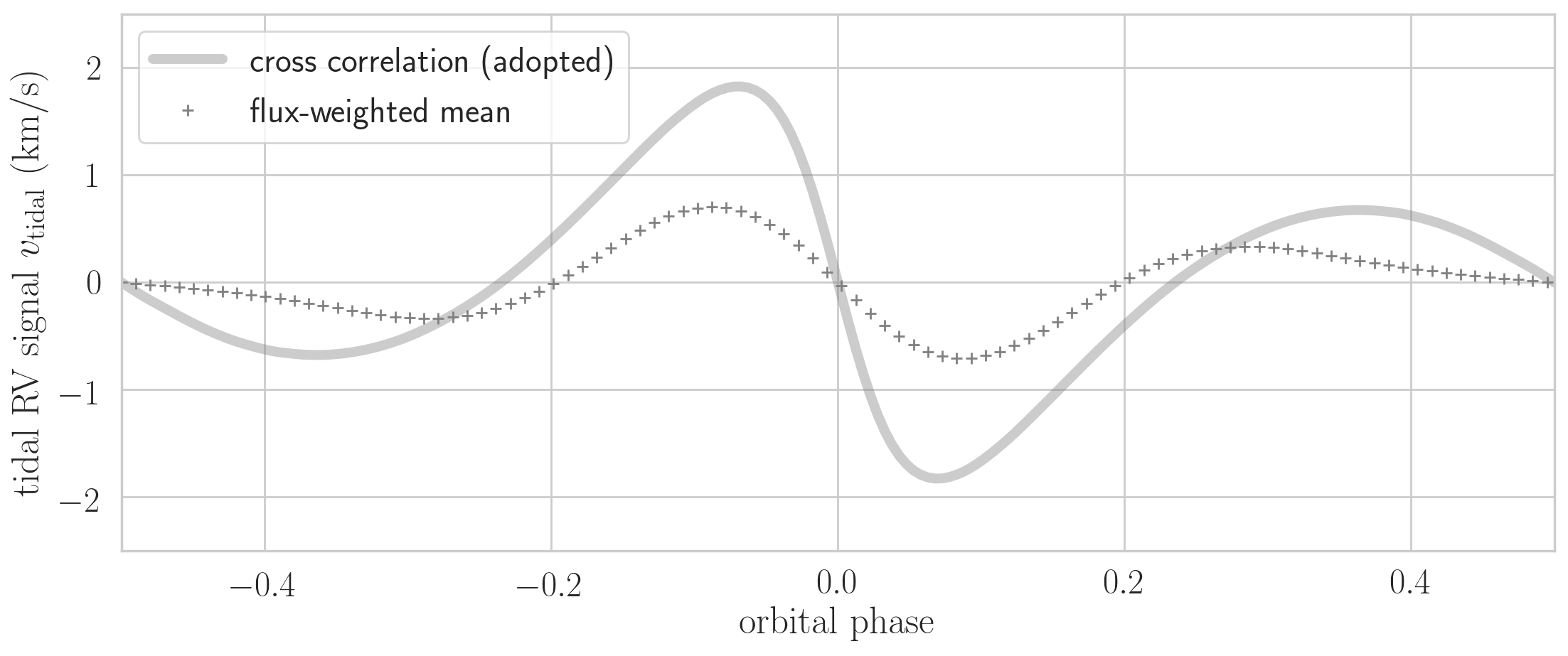}
    \caption{Tidal RV signal evaluated as the flux-weighted mean of the surface velocity field (crosses) has a smaller amplitude than the signal computed by modeling the cross-correlation function and by computing its peak (thick solid line). Note that the two models differ not only in terms of the amplitudes but also on the phases of the local maxima/minima and zero-crossings.}
    \label{fig:rvmethod}
\end{figure}

One simple method to evaluate tidal RV anomalies is to compute the mean of $V_j$ weighted by the flux $\Delta F_j$, $(\sum_j V_j \Delta F_j)/(\sum_j \Delta F_j)$. This is the prescription proposed in the seminal works by \citet{1941PNAS...27..168S} and \cite{1976ApJ...203..182W}, and has been adopted in many other works.
In reality, however, the RVs are derived via a more complicated procedure specific to each pipeline, and the simple ``flux-weighted mean velocity" has been shown to deviate from actual measurements in the case of the Rossiter--McLaughlin signal \citep{1924ApJ....60...15R, 1924ApJ....60...22M} originating from line-profile distortion due to transiting exoplanets \citep{2005ApJ...622.1118O, 2005ApJ...631.1215W, 2010ApJ...709..458H, 2011ApJ...742...69H}. 
This is also found to be the case in our current problem: Figure \ref{fig:rvmethod} compares the tidal RV signal evaluated as the flux-weighted mean (crosses) against the values from a model described below (thick solid line), for the same set of model parameters (mean of the posterior distribution) from our analysis in Section \ref{sec:results}. 
Because the RV data modeled in Section \ref{sec:results} were derived by computing cross correlation between the observed spectra and a synthetic template spectrum \citep{2012AN....333..663S}, we try to replicate the process as possible to model tidal RVs. The formulation here largely follows the one in \citet{2011ApJ...742...69H}.

We mainly consider the distortion of a single line at some specific wavelength and evaluate how this affects the RV values derived from a cross-correlation analysis. 
The line profile in velocity space $\mathcal{F}(v)$, in the presence of rigid rotation and macroturbulence, is given by the following convolution
\begin{equation}
    \mathcal{F}(v) = S(v) * M(v).
\end{equation}
Here
\begin{equation}
    M(v) = {{\sum_j \Theta_j(v-V_j) \Delta F_j}\over{\sum_j \Delta F_j}}
\end{equation}
is the broadening kernel, 
where
\begin{align}
    \label{eq:theta}
    \Theta_j(v) = {1\over2}\,\left[\norm\left(v; 0, {1\over2}\,\vmac^2\cos^2\gamma_j\right) + \norm\left(v; 0, {1\over2}\,\vmac^2\sin^2\gamma_j\right)\right],
    \quad 
    \norm(x; \mu, \sigma^2) \equiv {1 \over \sqrt{2\pi\sigma^2}}\,\exp\left[-{(x-\mu)^2 \over {2\sigma^2}}\right]
\end{align}
is the macroturbulence kernel for the radial-tangential model \citep{2005oasp.book.....G}. Here we assume equal contributions from the radial and tangential motions, and ignore the small Doppler shift due to flux difference between the rising and sinking gas streams (convective blueshift).\footnote{Another implicit assumption is that the absorption lines arise from the same equipotential surface on which we evaluated $\Delta F_j$. This is not exactly the case, but the difference has a negligible effect on the profile \citep{1998MNRAS.298..153S}.}
We assume that the intrinsic line profile $S(v)$ in velocity space is given by a Gaussian\footnote{Although the Vogit profile (convolution of Gaussian and Lorenzian) is physically more appropriate, the contribution from the Lorenzian part is minor here and this simplification is justified.}
\begin{equation}
    S(v) = \norm(v; 0, \beta_\mathrm{S}^2),
\end{equation}
where $\beta_\mathrm{S}^2=\beta^2_\mathrm{thermal} + \beta^2_\mathrm{mic} + \beta^2_\mathrm{IP}$ includes broadening contributions from the thermal motion, microturbulence, and  instrumental profile, respectively. 
We assume $\beta_\mathrm{thermal}=0.82\,\kms$ (corresponding to $T_\mathrm{eff}=4500\,\mathrm{K}$ and iron atoms), $\beta_\mathrm{mic}=1.0\,\kms$ \citep{2018AJ....156..125H, 2021arXiv210102212J}, and $\beta_\mathrm{IP}=2.31\,\kms$ (corresponding to the wavelength resolution $R=55,000$) for V723 Mon. These yield the total $\beta_\mathrm{S}=2.65\,\kms$. 
The resulting line profile is
\begin{equation}
    \mathcal{F}(v) \propto 
    \sum_j \left[\mathcal{N}\left(v-V_j; 0,\beta_\mathrm{S}^2 + {1\over 2}\,\zeta^2\cos^2\gamma_j\right) + 
    \mathcal{N}\left(v-V_j; 0, \beta_\mathrm{S}^2 + {1\over 2}\,\zeta^2\sin^2\gamma_j\right)\right]\,\Delta F_j.
\end{equation}

The cross correlation function (CCF) $C(v)$ is computed by convolving a template spectrum $T(v)$ against $\mathcal{F}(v)$. We assume that the template is a theoretical spectrum similar to the observed one \citep[as was the case in][]{2012AN....333..663S}, but without broadening due to instrumental profile: 
\begin{equation}
T(v)=\norm(v; 0, \beta_\mathrm{T}^2), \quad \beta_\mathrm{T}^2 \equiv \beta^2_\mathrm{thermal} + \beta^2_\mathrm{mic}.
\end{equation} 
Then the CCF is given by
\begin{equation}
    \label{eq:ccf}
    C(v) = (T \star \mathcal{F})(v) 
    \propto \sum_j \left[\mathcal{N}\left(v-V_j; 0, \beta_\mathrm{S}^2 + \beta_\mathrm{T}^2 + {1\over 2}\,\zeta^2\cos^2\gamma_j\right) + 
    \mathcal{N}\left(v-V_j; 0, \beta_\mathrm{S}^2 + \beta_\mathrm{T}^2 + {1\over 2}\,\zeta^2\sin^2\gamma_j\right)\right]\,\Delta F_j.
\end{equation}
Thus the shape of the CCF profile and the resulting RVs do not only depend on $V_j$ and $\Delta F_j$, but on the line-profile parameters $\zeta$ and $\beta \equiv \sqrt{\beta_\mathrm{S}^2 + \beta_\mathrm{T}^2}$.

The model RVs $\vtide$ are then derived as $\vtide = \mathrm{argmax}_v C(v)$. Given that the derivatives of $C(v)$ can be computed easily, $\vtide$ can often be found efficiently as the root of $\mathrm{d}C(v)/\mathrm{d}v=0$ using the Newton--Raphson method. However, we found that this method sometimes fails when the tidal deformation is large and the CCF has many points of inflection.\footnote{The CCF can even have two local maxima when the star is almost filling its Roche lobe, but this does not happen in our solution.} 
Thus we adopted a slower but more robust procedure: we first cast $\mathrm{d}C(v)/\mathrm{d}v=0$ into the form $v=f(v)$ and solve it iteratively starting from the flux-weighted mean of $V_j$ (which can be readily computed from the quantities in Section \ref{ssec:method_flux}), and use two Newton--Raphson steps to make the iterative solution converge efficiently to the best one. 

We have so far evaluated $\vtide$ by modeling the CCF for a single line at a particular wavelength, using the weights $\Delta F_j$ evaluated at single effective wavelength (see Section \ref{ssec:method_flux}). 
In reality, however, the CCF is computed from the spectrum with many absorption lines at different wavelengths, and so the actual CCF would be a weighted sum of the many CCFs computed in Equation~\ref{eq:ccf}. Assuming that $\beta$ and $\zeta$ are achromatic, this summation is equivalent to replacing $\Delta F_j$ in Equation~\ref{eq:ccf} with the value integrated over the wavelength range of the spectrum with a certain weight $W(\lambda)$. Since $\Delta F_j$ depends on the wavelength only through the limb- and gravity-darkening, this operation reduces to choosing the limb- and gravity-darkening coefficients $u_1$, $u_2$, and $y$ evaluated in the appropriate band.
The detailed shape of the weight $W(\lambda)$ is difficult to model, because it depends on the strengths and amounts of the lines 
as well as the \'{e}chelle orders used for RV extraction
that affect the wavelength region to which the CCF, and hence the derived RV, is most sensitive. We thus introduce this effective band as another parameter that may vary within a physically reasonable range, and take into account its uncertainty in evaluating the coefficients. See Section \ref{ssec:method_priors} for practical implementation of this model.

We note that the formulation presented here is not necessarily a unique one but needs to be adjusted depending on the exact procedure adopted to extract RVs. 
For example, RVs may be derived from features of the CCF other than the peak or by fitting a Gaussian to the CCF; the CCF may be computed using a binary mask rather than a theoretical template spectrum; or the RVs may be derived by directly fitting the observed spectra with a theoretical model.
Nevertheless the framework presented here remains useful for constructing similar models for RVs from different pipelines. 

\newcommand{\sjit}{\sigma_\mathrm{jitter}}

\subsection{Full RV Model, Likelihood, and Sampling}\label{ssec:method_full}

The RV measured at time $t_i$ was modeled as
\begin{equation}
    v(t_i) = -K \cos\left[2\pi\left({t_i-t_0} \over P\right)\right] + \gamma + \vtide(t_i),
\end{equation}
where $t_0$ is the time of the conjunction where the companion is in front of the red giant 
and $\gamma$ denotes the RV zero point.
The RV semi-amplitude $K$ is
\begin{equation}
    K = \left( 2\pi G \over P\right)^{1/3}\,{\mc \sin i \over (\mrg+\mc)^{2/3}},
\end{equation}
where $G$ is Newton's gravitational constant.

We assume that the measurement errors for RVs 
are independent and identical Gaussians with the variance of 
$\sigma_i^2+\sjit^2$. Here $\sigma_i$ is an internal error of the $i$th data point, and $\sjit$ models any other excess scatter that is not included in $\sigma_i$\footnote{
It is not uncommon that field red giant stars exhibit RV jitters of up to $\sim1\,\kms$ level \citep{2003AJ....125..293C}. Some of the presumably single giant stars observed by \citet{2012AN....333..663S} also show RV variations of $\mathcal{O}(0.1\,\kms)$.} 
and was inferred along with the other model parameters. 
Therefore the log-likelihood for a set of RV measurements $y_i$ at times $t_i$ is given by
\begin{equation}
    \ln \mathcal{L}
    = -{1\over 2} \sum_i 
    \left\{ 
    {\left[y_i-v(t_i)\right]^2 \over \sigma_i^2+\sjit^2} + \ln\left[2\pi(\sigma_i^2+\sjit^2)\right] \right\}.
\end{equation}

The whole code was implemented using {\tt JAX} \citep{jax2018github} and {\tt NumPyro} \citep{bingham2018pyro, phan2019composable}. We assumed the priors as described in Table \ref{tab:params} and Section \ref{ssec:method_priors}, and obtained posterior samples for the parameters using Hamiltonian Monte Carlo \citep{DUANE1987216, 2017arXiv170102434B}. We sampled until the resulting chains had the split $\hat{R}<1.01$ \citep{BB13945229} for all the parameters. 

\subsection{Priors}\label{ssec:method_priors}

We adopted the priors summarized in Table \ref{tab:params}. They are uninformative unless otherwise specified below. We note that these priors are independent from the information derived from ellipsoidal variations or eclipses of the Balmer lines.\\

\noindent
{\it RG mass $\mrg$ and radius $\rrg$} --- 
We adopt a prior uniform in $[0.5\,M_\odot, 3.0\,M_\odot]$ for the RG mass.
For the RG radius, we assume a Gaussian prior $\mathcal{N}(\rrg; 24.0\,R_\odot, 0.9\,R_\odot)$ based on the value derived from the SED, Gaia EDR3 distance \citep{2020arXiv201201533G}, and correction for non-stellar flux using measured dilution of the absorption lines \citep{2021arXiv210102212J}. 
Our definition of $R_\star$ (Section \ref{ssec:method_flux}) is not exactly the same as that in \citet{2021arXiv210102212J}, but the difference in the resulting solution is significantly smaller than the prior uncertainty and so can be ignored. 
\\

\noindent
{\it macroturbulence $\vmac$} --- The macroturbulence $\vmac$, along with rotation, shapes the rotation kernel (Equation \ref{eq:theta}) and can affect the RVs derived from CCFs. We adopt a Gaussian prior based on the relation from APOGEE DR13 \citep{2018AJ....156..125H}, which encompasses the values estimated by \citet{2021arXiv210102212J} from spectra and agrees with other measurements for RGB stars \citep[e.g.][]{2008AJ....135..892C}. The effect due to this uncertainty turns out to be minor, but would have been significant if the actual value was close to $v\sin i$ (see Section \ref{ssec:results_dep}). \\

\noindent
{\it profile width $\beta$} --- As detailed in Section \ref{ssec:method_rv}, this parameter represents the broadening of the CCF due to intrinsic widths of the absorption lines and that of the template. 
Larger values of $\beta$ tend to result in smaller $\vtide$ by smearing out the difference between pixels with different line-of-sight velocities $V_j$.
The estimate in Section \ref{ssec:method_rv} 
gives $\beta=2.95\,\kms$, 
but the actual value could be larger depending on the factors including exact broadening of the theoretical template used for the analysis, microturbulence, and wavelength dependence of the instrumental profile. Thus we adopt a half-normal prior centered on $2.95\,\kms$ with the width of $1\,\kms$. 
\\

\noindent
{\it limb- and gravity-darkening coefficients $u_1$, $u_2$, and $y$} --- 
Limb-darkening reduces the flux contribution from the surface elements with large line-of-sight velocities and reduces the amplitude of $\vtide$. Gravity darkening, on the other hand, enhances the amplitude because the effect decreases the flux from the underrepresented side in velocity space (e.g.~further reduces the ``red" flux in Figure \ref{fig:schematic}).
Their values in our model depend on the effective wavelength band defined through the CCF (see Section \ref{ssec:method_rv}). The effective band is difficult to quantify, but the following assumptions are reasonable: (i) the value is unlikely to be far from that obtained by integrating over the whole spectrum range with a uniform weight, because the lines relevant for RV measurements exist over the entire range, and (ii) the value should be bracketed by the values computed for the shortest and the longest wavelengths $(\lambda_\mathrm{min}, \lambda_\mathrm{max})$ in the spectrum.
We implement this prior knowledge as follows. We take $u'g'r'i'z'$-band coefficients theoretically computed with the ATLAS model \citep{2011A&A...529A..75C} for the effective temperature of $4500\,\mathrm{K}$, log surface gravity of $1.5$, and metallicity of $-1$ \citep{2021arXiv210102212J}, and interpolate them over the central wavelengths of the bands to obtain the coefficients $(u_1, u_2, y)$ as a function of wavelength. 
We then introduce a new parameter $\leff$ that represents the effective band. This parameter was sampled from a Gaussian with the central value of $(\lambda_\mathrm{min}+\lambda_\mathrm{max})/2$ and the width of $(\lambda_\mathrm{max}-\lambda_\mathrm{max})/4$.
Then we sample the coefficients from three independent Gaussians centered around $u_1(\leff)$, $u_2(\leff)$, and $y(\leff)$ computed using the above deterministic relations and widths of 0.1 to incorporate uncertainties in the theoretical calculations. This prior is insensitive to the adopted spectroscopic parameters within their uncertainties as evaluated by \citet{2021arXiv210102212J}.
\\

\noindent
{\it projected rotation velocity $v\sin i$} ---
Assuming tidal synchronization, our model automatically computes $v\sin i = 2\pi R_\star\sin i/P$. This has also been evaluated from several different sets of spectra \citep[see][]{2021arXiv210102212J}, but we did {\it not} include this information in the fit for two reasons. First, 
interpretation of the ``$v\sin i$" values of a tidally deformed star depends on how exactly they were extracted \citep{1998MNRAS.298..153S}.
Second, the measurements also depend on macroturbulence velocities adopted in those analyses, which are most likely different from each other and are not readily available.
Nevertheless, the value predicted from our model turned out to be in reasonable agreement with those existing measurements.

\section{Results}\label{sec:results}

We modeled RVs measured by \citet{2012AN....333..663S} from high-resolution ($R=55,000$) spectra obtained with the STELLA \'{e}chelle spectrograph on the 1.2~m STELLA-I telescope at the Teide Observatory \citep{2004AN....325..527S, 2008SPIE.7019E..0LW, 2010AdAst2010E..19S} between November 2006 and April 2010.\footnote{We also performed the same modeling for RVs from \citet{2014Obs...134..109G} with larger uncertainties and found a consistent result.} 
The spectra cover the wavelength range 388--882~nm and were reduced using the pipeline described in \citet{2008SPIE.7019E..0LW}. The RVs were determined from an order-by-order cross correlation analysis adopting a synthetic template spectrum \citep{1993sssp.book.....K} that roughly matches the target spectral classification.
Since \citet{2012AN....333..663S} initially identified the system as double-lined, 
the RVs were derived from the peak of the two-dimensional CCF \citep{2011A&A...531A..89W} as we exactly model here.
Our priors on $\beta$ and $\leff$ were chosen based on these information (Section \ref{ssec:method_priors}). 
The mean internal RV error is $\approx 0.2\,\kms$.
We removed one outlier at $\mathrm{BJD}=2454073.62965$ because the point was found to deviate from the model by more than $5\sigma$ and was not adequately modeled. 
We checked that the choice did not make a significant difference in the inferred parameters.

\begin{figure*}[htbp]
    \centering
    \epsscale{1.15}
    \plotone{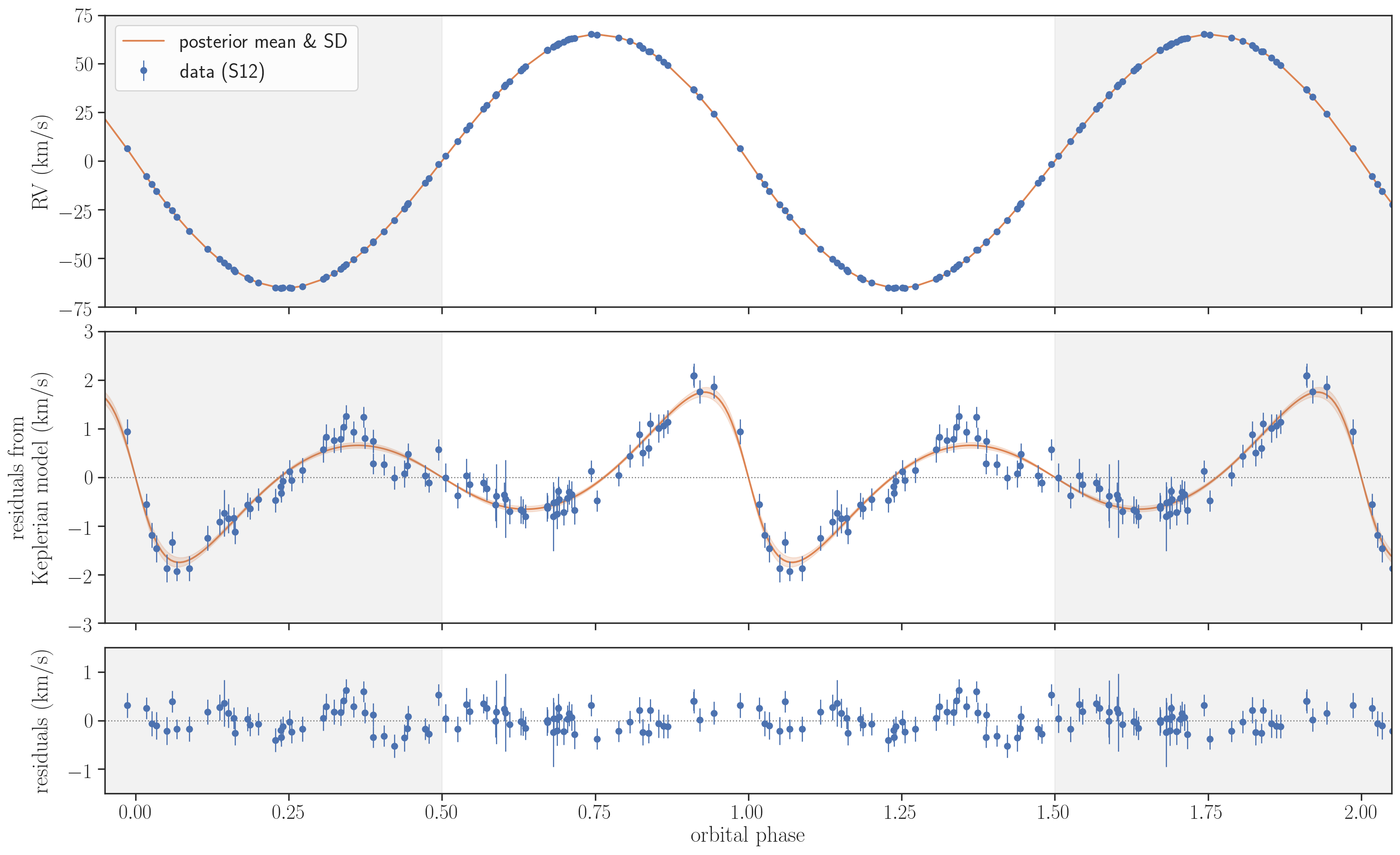}
    \caption{The observed and modeled RVs as a function of orbital phase. The blue filled circles are the RV data from \citet{2012AN....333..663S}. The orange solid line and the shaded region respectively show the mean and standard deviation of the models computed for posterior samples of the parameters. The gray-shaded region (phase less than 0.5 and larger than 1.5) shows the periodic repetition of the data and the model.
    {\it (Top)} --- RVs relative to the zero point $\gamma$. 
    {\it (Middle)} --- RVs relative to the Keplerian component plus $\gamma$.
    {\it (Bottom)} --- RVs relative to the full model.
    }
    \label{fig:fit}
\end{figure*}

The model based on the posterior samples of the parameters is compared with the data in Figure \ref{fig:fit}, and the resulting constraints on the parameters are summarized in Table \ref{tab:params} and Figure \ref{fig:corner}.
Our model successfully explains the periodic RV residuals from the Keplerian model, as shown in the middle and bottom panels of Figure \ref{fig:fit}. The shape and amplitude of the tidal RV signal constrain $\cos i$, $\mc/\mrg$, and $a/R_\star$, while the RV semi-amplitude pins down the mass function. Thus $\mc$ is determined from the RV signal and the prior on $\rrg$ alone.
The derived masses $\mc=2.95^{+0.17}_{-0.17}\,M_\odot$, $\mrg=0.82^{+0.13}_{-0.14}\,M_\odot$, and inclination $i=82.9^{+7.0}_{-3.3}\,\mathrm{deg}$ (medians and 68.3\% highest density intervals of the marginal posteriors) are all consistent with 
$\mc=3.04\pm0.06\,M_\odot$, $\mrg=1.00\pm0.07\,M_\odot$, and $i=87.0^{+1.7}_{-1.4}\,\mathrm{deg}$
derived by \citet{2021arXiv210102212J} using both RVs and ellipsoidal variations but without modeling the RV residuals.
The result provides additional, independent evidence for the companion mass and limits on the companion's luminosity derived by \citet{2021arXiv210102212J}, and eliminates the need for a third body as the origin of the non-Keplerian RVs.
Our larger error bars can partly be attributed to taking into account the uncertainties of the limb- and gravity-darkening coefficients 
and the RV scatter slightly larger than the internal error bars. 

\begin{deluxetable*}{l@{\hspace{.1cm}}cc@{\hspace{.15cm}}c}[!ht]
\tablecaption{System Parameters from our RV Modeling.\label{tab:params}}
\tablehead{
\colhead{} & \colhead{median \& $68.3\%$ HPDI} & \colhead{$90\%$ HPDI} & \colhead{prior} 
}
\startdata
red giant mass $\mrg$ ($M_\odot$) & 
$0.82^{+0.13}_{-0.14}$ & $[0.62, 1.07]$
& $\u(0.5, 3.0)$\\
red giant radius $\rrg$ ($R_\odot$) & 
$24.25^{+0.88}_{-0.89}$ & $[22.74, 25.67]$
& $\norm(24.0, 0.9, 15)$
\\
mass ratio $\mc/\mrg$ & 
$3.58^{+0.38}_{-0.54}$ & $[2.85, 4.37]$
& $\logu (\exp(0), \exp(3))$\\
companion mass $\mc$ ($M_\odot$) & 
$2.95^{+0.17}_{-0.17}$ & $[2.68, 3.23]$
& \nodata \\
semi-major axis over red giant radius $a/\rrg$ & 
$4.142^{+0.091}_{-0.093}$ & $[4.001, 4.301]$
& \nodata\\
RV semi-amplitude $K$ ($\kms$) & 
$65.268^{+0.061}_{-0.052}$ & $[65.17, 65.36]$
& \nodata\\
binary mass function ($M_\odot$) & 
$1.7264^{+0.0049}_{-0.0041}$ & $[1.7188, 1.7337]$
& \nodata\\
time of conjunction $t_0$ 
($\mathrm{BJD-2450000}$)
& 
$5575.0653^{+0.0073}_{-0.0071}$ & $[5575.0533, 5575.0767]$
& $\u(5574.5954, 5575.5954)$\\
orbital period $P$ (days) & 
$59.9376^{+0.0009}_{-0.0010}$ & $[59.9359, 59.9392]$
& $\norm(60, 1, 58)$\\
cosine of orbital inclination $\cos i$  & 
$0.12^{+0.06}_{-0.12}$ & $[0.00, 0.28]$
& $\u(0,1)$\\
profile width $\beta$ ($\kms$) & 
$3.93^{+0.42}_{-0.90}$ & $[2.96, 4.99]$
& $\norm(2.95, 1, 2.95)$\\
macroturbulence velocity $\vmac$ ($\kms$) & 
$5.52^{+0.97}_{-1.36}$ & $[3.78, 7.48]$
& $\norm(5.3, 1, 1)$\\
effective wavelength $\leff$ ($\mathrm{nm}$) & 
$626^{+100}_{-112}$ & $[475, 803]$
& $\norm(635.0, 123.5, 388)$ \\
limb-darkening coefficient $u_1$ & 
$0.64^{+0.15}_{-0.20}$ & $[0.37, 0.94]$
& \nodata\\
limb-darkening coefficient $u_2$ & 
$0.17^{+0.13}_{-0.12}$ & $[-0.04, 0.39]$
& \nodata\\
gravity-darkening coefficient $y$ & 
$0.46^{+0.09}_{-0.10}$ & $[0.31, 0.64]$
& \nodata\\
RV zero point $\gamma$ ($\kms$) & 
$1.885^{+0.031}_{-0.030}$ & $[1.835, 1.937]$
& $\u(-10,10)$\\
RV jitter $\sjit$ ($\kms$) & 
$0.168^{+0.026}_{-0.030}$ & $[0.121, 0.214]$
& $\logu (\exp(-5), \exp(0))$\\
projected rotation velocity $v\sin i$ ($\kms$) & 
$20.18^{+0.76}_{-0.88}$ & $[18.80, 21.52]$
& \nodata \\
\enddata
\tablecomments{Values listed here report the medians and $68.3\%$/$90\%$ highest posterior density intervals (HPDIs) of the marginal posteriors, which were found to be unimodel for all the parameters. Priors --- $\mathcal{N}(\mu, \sigma^2, l)$ means the normal distribution centered on $\mu$ and with variance $\sigma^2$; when $l$ is specified the normal distribution is truncated at the lower limit $l$. $\mathcal{U} (a,b)$ and $\mathcal{U}_{\ln} (a,b)$ are the uniform- and log-uniform probability density functions between $a$ and $b$, respectively. Dots indicate the parameters that were computed from the samples of the ``fitted" parameters whose priors were explicitly specified.
}
\end{deluxetable*}

We note that the derived RG mass is sensitive to the adopted prior on the RG radius, while the companion mass is less so, as was also noted by \citet{2021arXiv210102212J}. 
This is because the tidal RVs (as well as ellipsoidal variations) constrain $\cos i$ and the degree of tidal deformation $(\mc/\mrg)(\rrg/a)^3$, and so the decrease in $\rrg$ must be compensated by a larger $\mc/\mrg$, and hence smaller $\mrg$ for the fixed mass function. This positive correlation between $\mrg$ and $\rrg$ is seen in Figure \ref{fig:corner}.
If we instead adopt $\rrg=22.2\pm0.8\,R_\odot$ from the SED modeling without veiling correction by \citet{2021arXiv210102212J}, we find $\mrg=0.63^{+0.06}_{-0.12}\,M_\odot$, $\rrg=22.55^{+0.69}_{-0.75}\,R_\odot$, and $\mc=2.74^{+0.10}_{-0.19}\,M_\odot$.
The change in the companion mass $\mc$ is smaller than that in $\mrg$ because of the anti-correlation between $\mrg$ and $\mc/\mrg$ as described above. Thus the conclusion that the companion has $\approx 3\,M_\odot$ is robust against the uncertainty in the RG mass.

\begin{figure*}[htbp]
    \centering
    \epsscale{1.15}
    \plotone{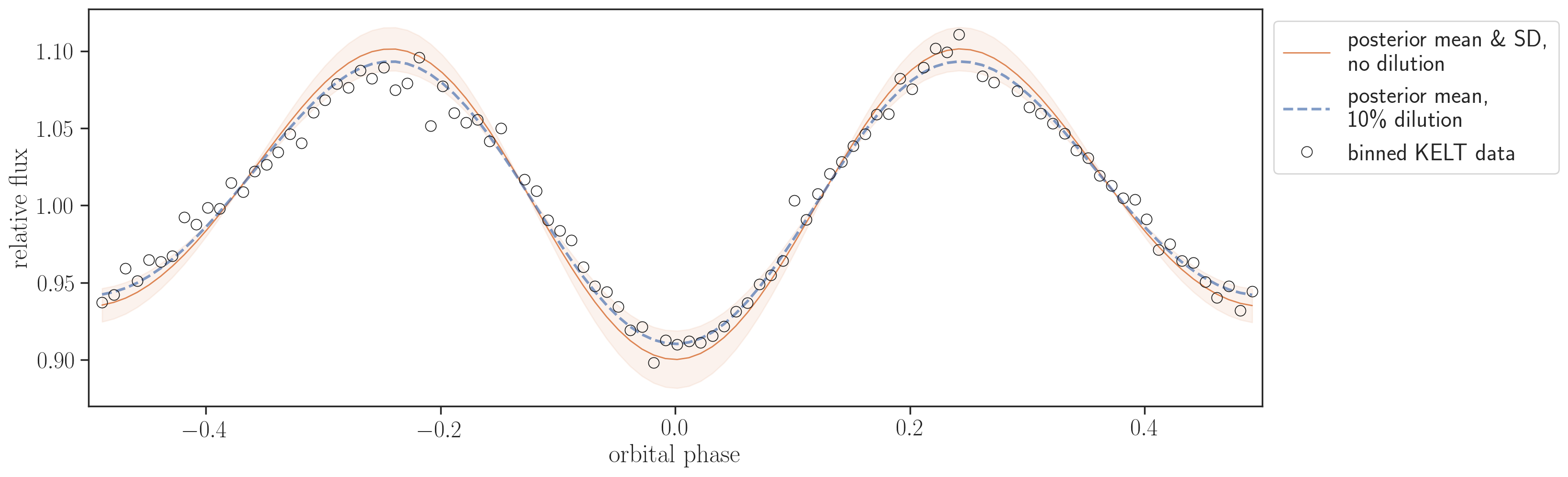}
    \caption{The flux variation {\it predicted} from our RV modeling compared with the KELT light curve. The data were phase-folded using the mean ephemeris derived from the RV modeling and averaged into 100 bins.}
    \label{fig:lc}
\end{figure*}

In Figure \ref{fig:lc}, we compare the light curve {\it predicted} from our RV model (computed as $\sum_j \Delta F_j$) with the data from the Kilodegree Extremely Little Telescope \citep[KELT;][]{2007PASP..119..923P}. The data were retrieved from the NASA Exoplanet Archive,\footnote{\url{https://exoplanetarchive.ipac.caltech.edu/}} phase-folded using the mean ephemeris derived from the RV modeling, and averaged into 100 bins.
In computing the light curve models,
the limb- and gravity-darkening coefficients derived from the RV fit were replaced with the values computed \citep{2011A&A...529A..75C} for the $R$-band, which is similar to the KELT band pass \citep{2007PASP..119..923P}. 
We show two sets of predictions: the orange solid line and shaded region show the mean and standard deviation of the posterior models assuming no dilution, respectively, and the blue dashed line shows the mean posterior model assuming 10\% dilution relative to the RG flux (not the total flux) due to the veiling effect, where normalization of each model is adjusted to best match the data.
Figure \ref{fig:lc} shows that the data barely match the prediction of the zero-dilution model, and that the agreement is better for the model with $10\%$ dilution.
Although \citet{2021arXiv210102212J} assumed no dilution in their analysis of the KELT light curve, the $\sim10\%$ dilution favored by our RV model appears to be reasonable given the line dilution analysis by \citet{2021arXiv210102212J} (their Figure 8 left) and the wide effective width ($318\,\mathrm{nm}$) of the KELT band \citep{2007PASP..119..923P}.
Thus we conclude that our RV model is consistent with the observed ellipsoidal variations within the uncertainty of the veiling flux and RG radius. This comparison illustrates the importance of the tidal RV signal as an independent means to check any flux contamination in the light curve.

\clearpage

\subsection{Sensitivity to the Line-profile Parameters} \label{ssec:results_dep}

\begin{figure*}
    \centering
    \epsscale{0.55}
    \gridline{
          \fig{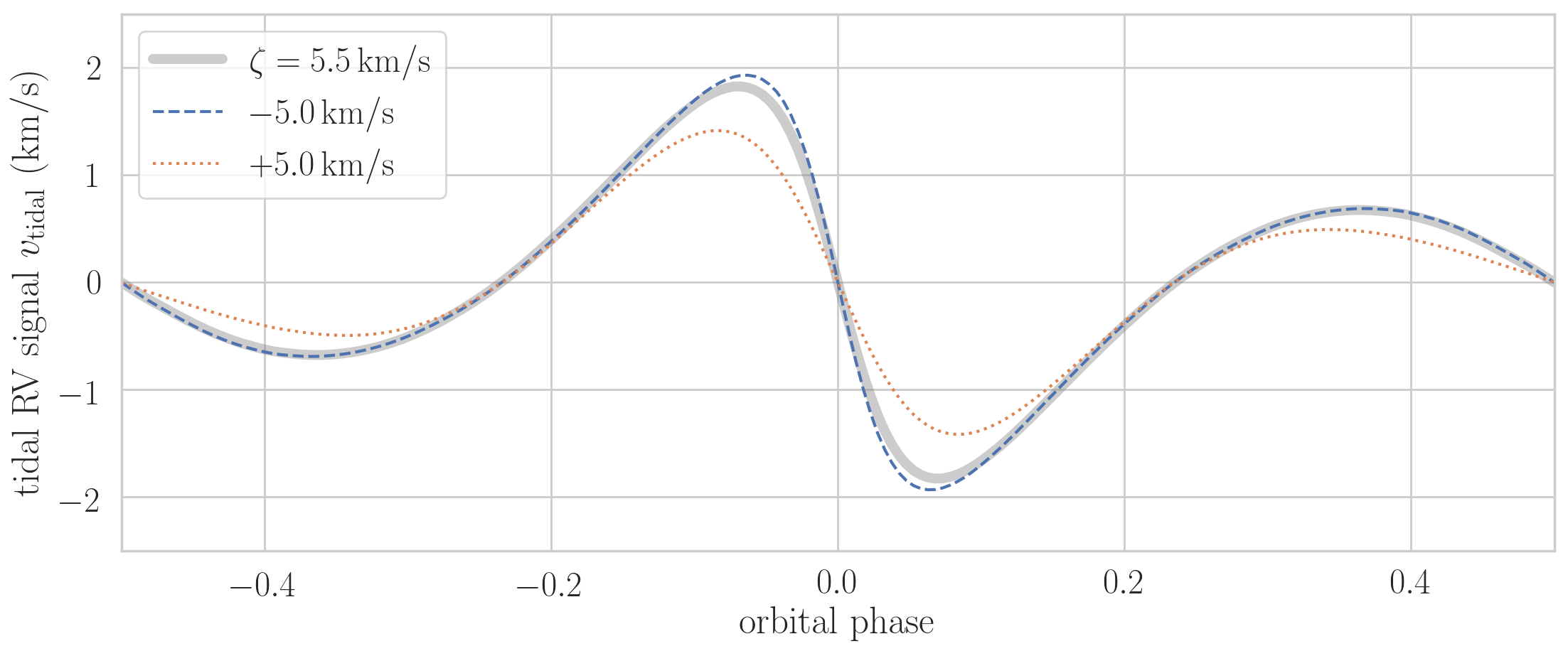}{0.5\textwidth}{(a) macroturbulence $\zeta$}
          \fig{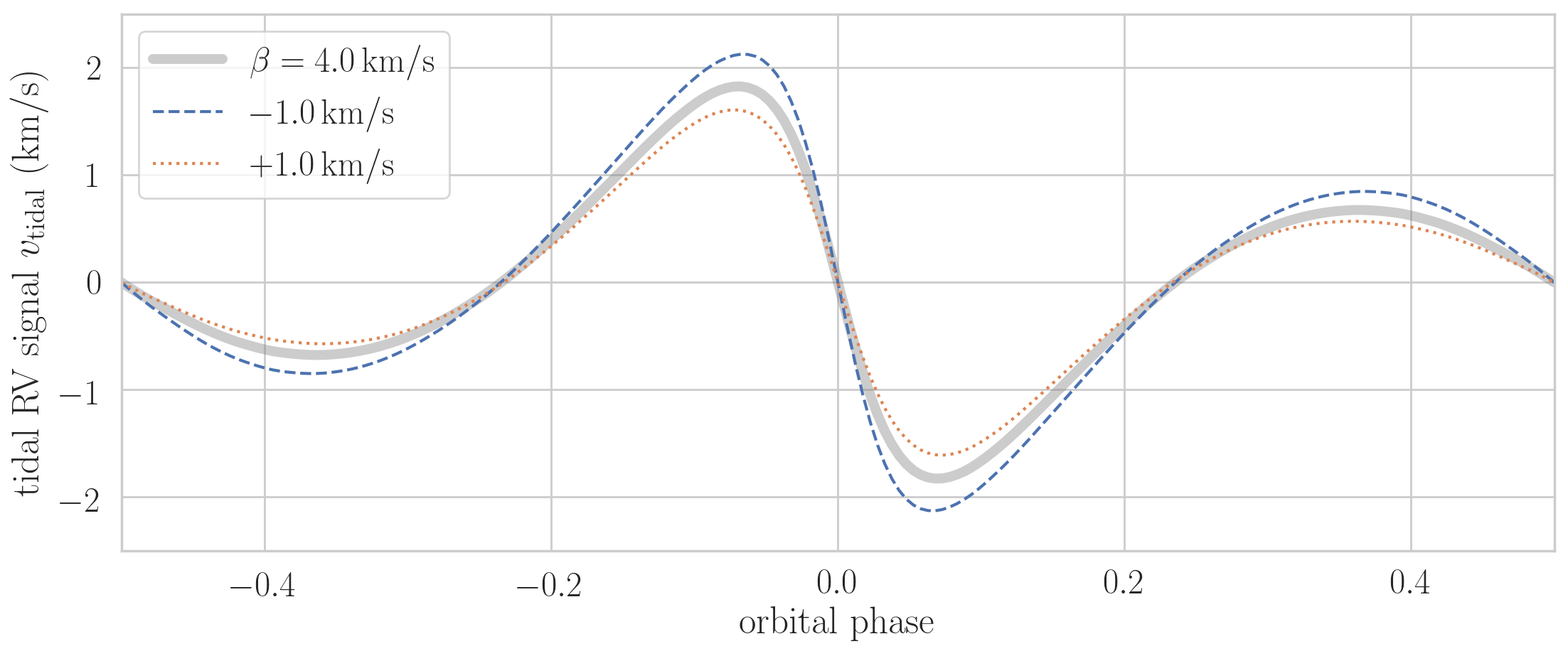}{0.5\textwidth}{(b) profile width $\beta$}
         }
    \caption{Dependence of the tidal RV signal on (a) macroturbulence $\zeta$ and (b) profile width $\beta$.}
    \label{fig:dependence}
\end{figure*}

Our RV model includes two additional parameters, macroturbulence $\zeta$ and profile width $\beta$, which are not required when the tidal RV is modeled as the flux-weighted mean velocity \citep{1976ApJ...203..182W}. 
These parameters are not constrained by the RV data but mostly determined by the adopted priors (see Table \ref{tab:params}). Although we believe they are reasonable \citep[and $\zeta$ is constrained from the spectra;][]{2021arXiv210102212J}, we show in Figure \ref{fig:dependence} how the model $\vtide$ depends on these parameters to gauge their potential impacts on the other inferred parameters. Here the thick gray lines 
show the model computed for the mean parameter values of the posterior distribution,
and the dashed and dotted lines show models where each parameter is perturbed by the values shown in the legends.
The results show that it is essential to take into account their uncertainties as we did. In particular, the prior on the macroturbulence parameter needs to be chosen carefully. When the value is a significant fraction of $v\sin i$, as is the case for the dotted curve corresponding to $\zeta\approx 11\,\kms$, the amplitude of the tidal RV depends significantly on this parameter. This can indeed be the case for some other giant stars.

\subsection{Sensitivity to the Limb-Darkening Profile}
\label{ssec:results_ld}

The model atmospheres adopting spherical geometry suggest that the limb of giant stars with low surface gravity can be substantially darker than predicted by simple parametric laws \citep[see, e.g., Figure 1 of ][]{2000A&A...364..265O} as we have adopted in the above analysis. 
To check on the sensitivity of our analysis to the limb-darkening profile, we redid the fit replacing the quadratic profile with the ones based on the intensity calculations from the PHOENIX model atmospheres with spherical geometry \citep{2013A&A...553A...6H}.\footnote{\url{http://phoenix.astro.physik.uni-goettingen.de}} We used a model computed for the same atmospheric parameters as adopted in Section \ref{ssec:method_priors}, retrieved the intensity values for $78$ different $\cos \gamma$ and for the wavelengths spanning $352.5$--$952.5\,\mathrm{nm}$ at $100\,\mathrm{nm}$ intervals, and linearly interpolated them to compute $I(\cos\gamma, \leff)$ in our RV model. We also incorporated $10\%$ fractional uncertainty in the limb-darkening profile to take into account uncertainties in the model and in the adopted atmospheric parameters. For this analysis, we adopted 3072 {\tt HEALPix} pixels to ensure numerical stability for the updated limb-darkening profile with a sudden intensity drop at the limb.
We found the results consistent with the above analysis using the quadratic law, including $\mc=2.97^{+0.15}_{-0.17}\,M_\odot$, $i=82.8^{+7.2}_{-3.3}\,\mathrm{deg}$, and $\mrg=0.83^{+0.11}_{-0.15}\,M_\odot$. Thus we conclude that our result is robust against the uncertainty of the limb-darkening profile. 
This appears reasonable given that the deviation from the quadratic law occurs mainly at $\cos\gamma\lesssim 0.25$, where the PHOENIX atmospheres predict almost zero intensities. The severe darkening results in the loss of stellar flux in the outermost $\approx 1-\sqrt{1-0.25^2}=3\%$ of the stellar disk. This is equivalent to a $<1\,\kms$ change in $v\sin i$ and so plays a minor role in shaping the line profile. 
On the other hand, we found a slightly larger amplitude for the tidal RVs computed as the flux-weighted mean when the profile from the PHOENIX model was adopted.



\section{Summary and Discussion} \label{sec:summary}

We showed that the periodic RV residuals of V723 Mon are quantitatively explained by a model incorporating tidal deformation of the RG star and associated distortion of the absorption line profile. Our RV modeling constrains the companion mass to be
$M_\bullet = 2.95\pm0.17\,M_\odot$ and orbital inclination to be $i=82.9^{+7.0}_{-3.3}\,\mathrm{deg}$.
This provides additional evidence for the low-mass black hole companion in the mass gap as inferred by \citet{2021arXiv210102212J}, and eliminates the need for a third body to explain the periodic RV residuals. The derived inclination indicates that the companion should be eclipsed by the red giant, and thus also supports the limits on the companion's luminosity based on the absence of eclipses \citep{2021arXiv210102212J}.
Importantly, the constraint is independent from ellipsoidal variations or the eclipses of Balmer emission, which both include signals of unclear physical origin. Indeed, our RV modeling mildly favors $\sim10\%$ flux dilution in the KELT band that was not taken into account in the analysis by \citet{2021arXiv210102212J}. 
This illustrates an advantage of the tidal RV signal as a means to measure component masses in tidally interacting, single-lined binaries: any contaminating non-stellar flux, as long as its spectrum is continuous, does not significantly affect the positions and shapes of the absorption lines from which RVs are measured. 

The same signal will be useful 
for ``dynamical" mass measurements in other non-eclipsing post main-sequence binaries,
including the ones where the companion is not a black hole, without photometric light curves. 
The amplitude of $\vtide$ is of order $(v\sin i)\cdot q_\mathrm{comp} \cdot (R_\star/a)^3$, where $q_\mathrm{comp}$ is the companion mass relative to the star for which RVs are measured. Thus for a synchronized and circularized binary, its amplitude $K_\mathrm{tidal}$ relative to the semi-amplitude of the orbital RV $K$ is simply given by
\begin{equation}
    {K_\mathrm{tidal} \over K}
    \sim \left(R_\star \over a \right)^4(1+q_\mathrm{comp}),
\end{equation}
which is $\mathcal{O}(1\%)$ for V723 Mon. This estimate suggests that the amplitude of $\vtide$ can well reach $\mathcal{O}(100\,\mathrm{m\,s^{-1}})$ even in less extreme systems than V723 Mon. This is not the precision usually required for binary studies, but precisions better than this are routinely achieved in Doppler searches for exoplanets. This work motivates such high-precision RV measurements for binaries exhibiting strong tidal interactions.
We also echo the original note by \citet{1941PNAS...27..168S} that
the tidal signal could be relevant for interpreting the eccentricities of such binaries when their precise values matter, e.g. for studying details of tidal orbital circularization \citep[e.g.][]{1995A&A...296..709V, 2018ApJ...867....5P} or for precise evaluation of the apsidal precession rate \citep[e.g.][]{2017AJ....154....4P}. 

The mass measurement with the tidal RVs will be especially valuable for non-eclipsing systems where the ellipsoidal variations are swamped or contaminated by other light sources, including quasi-periodic modulations due to active regions (spots) that have a similar period to the orbital one in synchronized binaries. Although the presence of spots may also make it challenging to measure precise RVs, line-profile distortion by spots is usually localized in velocity space, and so could still be distinguished from the global distortion by tides via a careful analysis of the CCF shapes. Even in the absence of spots, direct modeling of the line-profile variations, as has been proposed by \citet{1998MNRAS.298..153S},
in principle provides more information and is less model dependent than modeling RV time series alone as we have done. Such an analysis is beyond the scope of this paper.

\acknowledgements

The authors thank Chris Kochanek for useful comments on the early manuscript of the paper, and the anonymous referee for an important note on the limb-darkening profile.
The authors are also grateful to Todd Thompson, Kris Stanek, Tharindu Jayasinghe, Klaus G. Strassmeier, and Michael Weber for sharing the STELLA RV data as well as the information on the relevant references. 
KM thanks Hajime Kawahara for introducing {\tt JAX} and {\tt NumPyro} and for sharing computational resources.
Work by TH was supported by JSPS KAKENHI Grant Number JP19K14783.

\software{
corner \citep{corner}, HEALPix \citep{2005ApJ...622..759G}, healpy \citep{Zonca2019}, JAX \citep{jax2018github}, NumPyro \citep{bingham2018pyro, phan2019composable}
}

\appendix 

\begin{figure*}[htbp]
    \centering
    \epsscale{1.15}
    \plotone{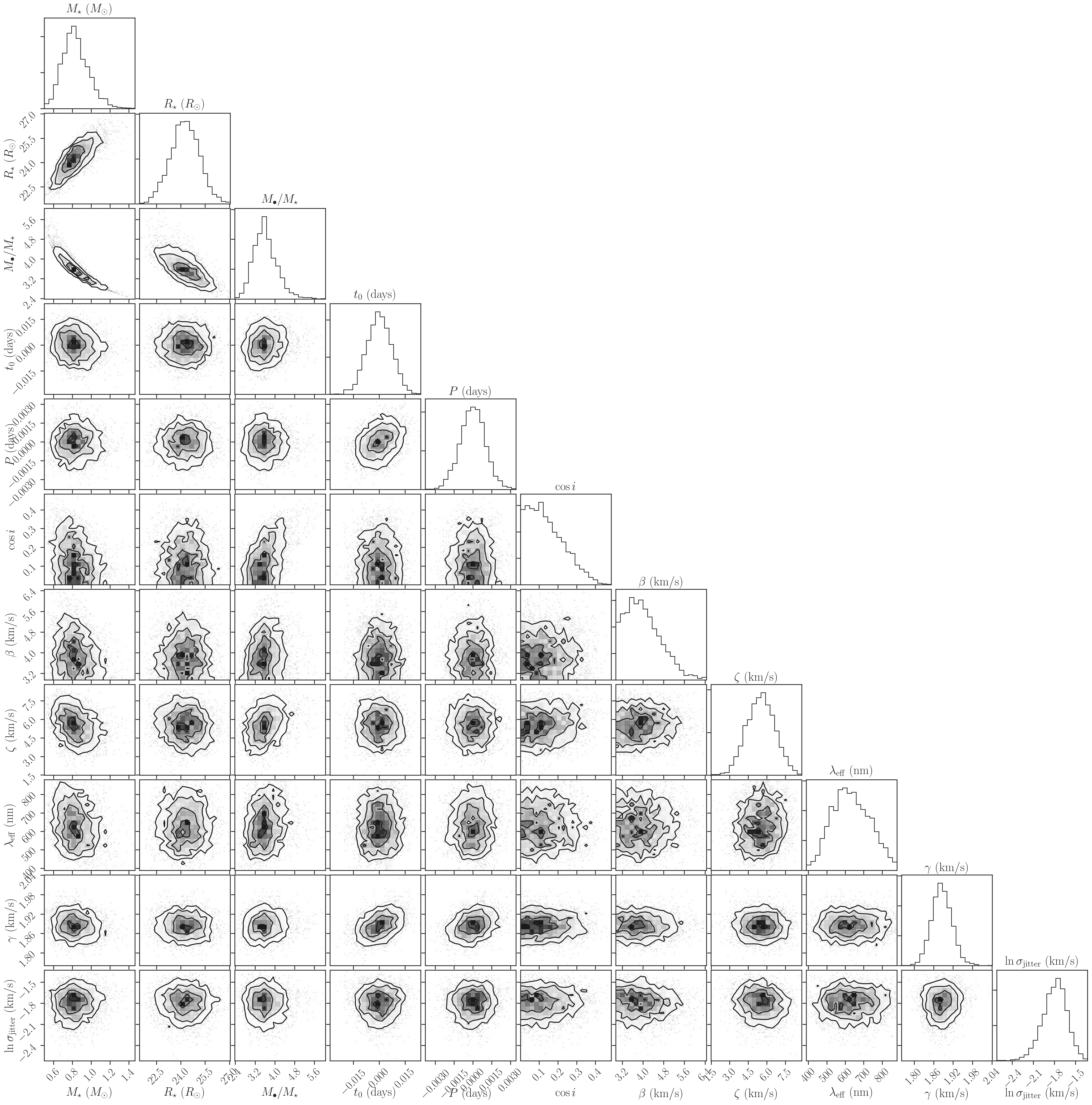}
    \caption{Corner plot \citep{corner} created from the posterior samples of the model parameters in Section \ref{sec:results}. For $t_0$ and $P$, values relative to their medians are shown for clarity.} 
    \label{fig:corner}
\end{figure*}

Figure \ref{fig:corner} shows one- and two-dimensional histograms of the posterior samples of the model parameters (Section \ref{sec:results}) to visualize their correlations. 

\bibliographystyle{aasjournal}



\end{document}